# In-situ study of rules of nanostructure evolution, severe plastic deformations, and friction under high pressure

*Feng Lin*[1], Valery I. Levitas*[1,2], Krishan K. Pandey[3], Sorb Yesudhas[1], and Changyong Park[4]*

[1]Department of Aerospace Engineering, Iowa State University, Ames, Iowa 50011, USA

[2]Department of Mechanical Engineering, Iowa State University, Ames, Iowa 50011, USA

[3]High Pressure & Synchrotron Radiation Physics Division, Bhabha Atomic Research Centre, Mumbai 400085, India

[4]HPCAT, X-ray Science Division, Argonne National Laboratory, Argonne, Illinois 60439, USA

Feng Lin  flin1@iastate.edu

Valery Levitas  vlevitas@iastate.edu

Krishan Pandey  kkpandey@barc.gov.in

Sorb Yesudhas  sorbya@iastate.edu

Changyong Park  cypark@anl.gov

**Abstract**

Severe plastic deformations under high pressure are used to produce nanostructured materials but were studied ex-situ. We introduce rough diamond anvils to reach maximum friction equal to yield strength in shear and perform the first in-situ study of the evolution of the pressure-dependent yield strength and nanostructural parameters for severely pre-deformed Zr. ω-Zr behaves like perfectly plastic, isotropic, and strain-path-independent. This is related to reaching steady values of the crystallite size and dislocation density, which are pressure-, strain- and strain-path-independent. However, steady states for α-Zr obtained with smooth and rough anvils are different, which causes major challenge in plasticity theory.

Impact statement

In-situ study of severe plastic deformation of ω-Zr with rough diamond anvils revealed that pressure-dependent yield strength, crystallite size, and dislocation density are getting steady and plastic strain- and strain-path-independent.

## 1. Introduction

Processes involving severe plastic deformations (SPD) under high pressure are common in producing nanostructured materials [1-8], in functional materials experiencing extreme stresses under contact friction, collision, and penetration, and in geophysics [9,10]. The effects of SPD under high pressure on microstructure evolution are mostly studied with high-pressure torsion (HPT) with metallic or ceramic anvils [1-4]. Stationary states after SPD in terms of torque, hardness, grain size, and dislocation density are well-known in literature, particularly after HPT, along with many cases where they were not observed [1-8, 11]. However, all these results *were not observed in-situ but obtained postmortem* after pressure release and further treatment during sample preparation for mechanical and structural studies (see supplement). The direct effect of pressure and the combined effect of pressure and plastic straining on the yield strength, crystallite size, and dislocation density were not determined. This is very important because, as we will see, the yield strength of the ω-

Zr doubles at ~13 GPa, but hardness and, consequently, yield strength after pressure release are independent of the pressure at HPT [12]. During pressure release after HPT of Ni, crystallite size increases, and dislocation density decreases by a factor of 2 [13]. Similar results were obtained for Zr under hydrostatic loading [14].

Robust method for measurement of the yield strength in compression $\sigma_y(p)$ under high pressure $p$ is lacking. The main difficulty in studying plasticity, structural changes, and contact friction is that they depend on five components of the plastic strain tensor $\boldsymbol{\varepsilon}_p$ and its entire path $\boldsymbol{\varepsilon}_p^{path}$, making an unspecifiable number of combinations of independent parameters. The yield surface in the 5D deviatoric stress $\boldsymbol{s}$ space $f(\boldsymbol{s}, \boldsymbol{\varepsilon}_p, \boldsymbol{\varepsilon}_p^{path}) = \sigma_y(p)$ depends on $p$, $\boldsymbol{\varepsilon}_p$, and $\boldsymbol{\varepsilon}_p^{path}$, demonstrating strain hardening/softening and strain-induced anisotropy. This complexity makes it impossible to determine the complete evolution of the yield surface, even at small strains and ambient condition. For measurement of yield strength at high pressure, all methods [15-17] treat the yield surface as $f(\boldsymbol{s}) = \sigma_y(p)$, i.e., like for perfectly plastic material (for which the yield surface is independent of $\boldsymbol{\varepsilon}_p$ and $\boldsymbol{\varepsilon}_p^{path}$, i.e., is fixed in the 5D stress space), and dependence on $\boldsymbol{\varepsilon}_p$ and $\boldsymbol{\varepsilon}_p^{path}$ is neglected and merged in pressure, which causes large error. One of the methods to determine the yield strength in shear $\tau_y = \sigma_y/\sqrt{3}$ in diamond anvil cell (DAC) is based on applying the simplified equilibrium equation $\frac{d\bar{P}}{dr} = -\frac{2\tau_f}{h}$, assuming the anvil-sample contact friction stress $\tau_f = \tau_y$ [16-18] (see supplement). Here, $\bar{P}$ is the pressure averaged over the sample thickness $h$. However, recent experiments [15, 19] show that $\tau_f < \tau_y$. Coupled simulations and experiments demonstrate that $\tau_f = \tau_y$ only in a small region, even above 100 GPa [20]. We introduce *rough diamond anvil (rough-DA)*, whose culet is roughly polished to increase friction (Figure 1). We demonstrated that maximum friction $\tau_f = \tau_y$ is reached for rough-DA, which allowed us to robustly determine $\sigma_y(p)$ and plastic friction.

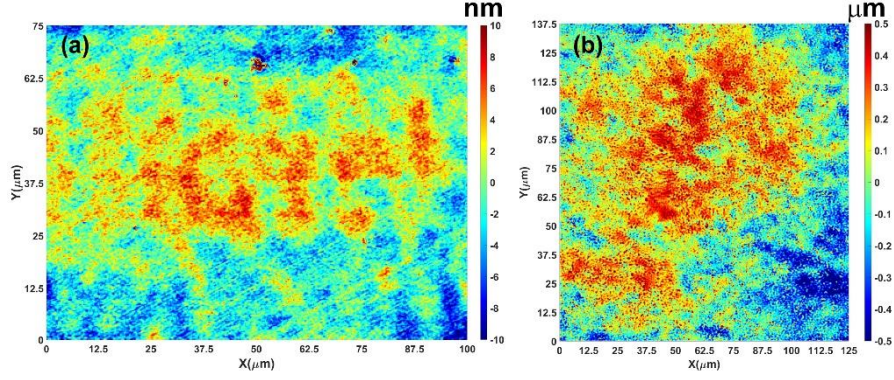

**Figure 1. Surface asperity profile of a smooth anvil and a rough-DA.** (a) Traditional smooth-DA with an asperity profile range [-10 nm; 10 nm] and (b) rough-DA with range [-500 nm; 500 nm].

It was hypothesized in [18] that, above some level of plastic strain in monotonous straining (straining path without sharp changes in directions), the initially isotropic polycrystalline materials deform as perfectly plastic and isotropic with a strain path-independent surface of the perfect plasticity $\varphi(\boldsymbol{s}) = \sigma_y(p)$ (Figure 2). Some qualitative supportive arguments for the perfect plastic behavior are presented in [18], but the quantitative experimental proof is lacking for any material. Here, we severely pre-deformed commercial Zr by multiple rolling until saturation of its hardness. We show that after the α-ω phase transformation, for four different compression stages (i.e., for very different $\boldsymbol{\varepsilon}_p$ and $\boldsymbol{\varepsilon}_p^{path}$), all pressure distributions of ω-Zr are described by single function $\sigma_y = 1.24 + (0.0965 \pm 0.0016)p \ (GPa)$. This is possible only if the *material behaves like perfectly plastic, isotropic, and independent of $\boldsymbol{\varepsilon}_p$ and $\boldsymbol{\varepsilon}_p^{path}$*. Similarly, friction stress $\tau_f = \tau_y = 0.72 + (0.0557 \pm 0.0009)p \ (GPa)$ is also *independent of $\boldsymbol{\varepsilon}_p$ and $\boldsymbol{\varepsilon}_p^{path}$*. The perfectly plastic state is connected to reaching a steady nanostructure, determined here by in-situ synchrotron XRD in terms of crystallite (grain) size $d$ and dislocation density $\rho$, which do not change under successive plastic straining. For rough-DA in α-Zr at the beginning of α-ω transformation, $d_\alpha$ is smaller, and $\rho_\alpha$ is larger than those from smooth anvils, i.e., *rough-DA produces a different, more refined steady*

*nanostructure*. The steady nanostructure for ω-Zr after transformation is the same for smooth and rough-DAs and is pressure-independent.

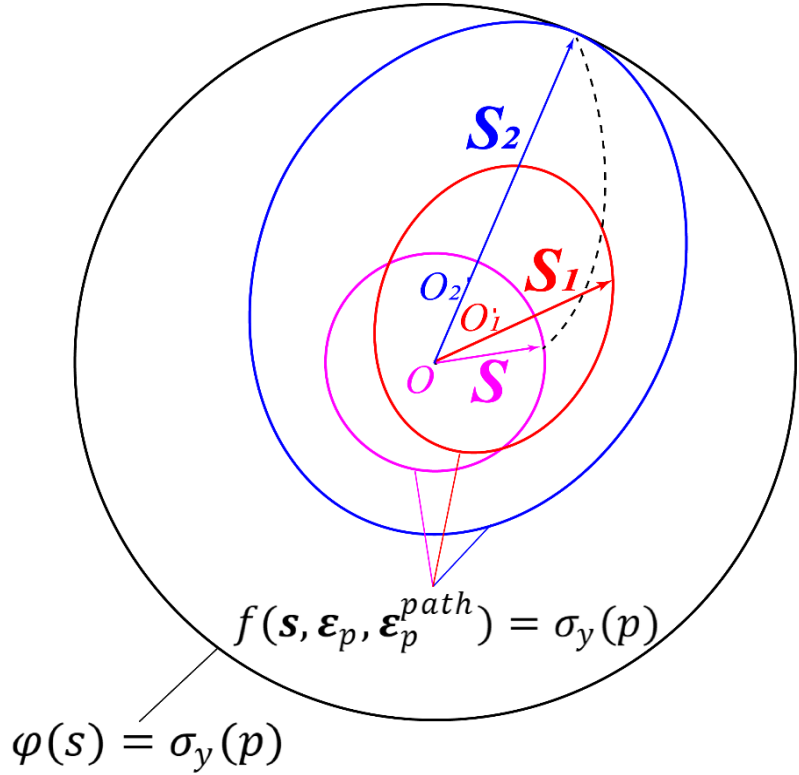

**Figure 2. Evolving yield surface and fixed surface of perfect plasticity.** Schematic of the evolution of the yield surface $f(\boldsymbol{s}, \boldsymbol{\varepsilon}_p, \boldsymbol{\varepsilon}_p^{path}) = \sigma_y(p)$ until it reaches the fixed surface of perfect plasticity $\varphi(\boldsymbol{s}) = \sigma_y(p)$ in "5D" space of deviatoric stresses $\boldsymbol{s}$ at fixed $p$. The initial yield surface and $\varphi(\boldsymbol{s}) = \sigma_y(p)$ are isotropic (circles). Two other yield surfaces depend on $\boldsymbol{\varepsilon}_p$ and $\boldsymbol{\varepsilon}_p^{path}$, and acquire strain-induced anisotropy, namely shifted centers $O_1$ and $O_2$ (back stress) and ellipsoidal shape due to texture. When the yield surface reaches $\varphi(\boldsymbol{s}) = \sigma_y(p)$, the material deforms like perfectly plastic, isotropic with the fixed surface of perfect plasticity.

## 2. Materials and methods

We heavily pre-deformed the commercially pure (99.8%) α-Zr slab with an initial thickness of 5.25 mm by multiple rolling down to 163-165 μm until saturation of its hardness. 3 mm diameter disks were punched out for compression in DAC with rough-DAs, and smooth-DAs for comparison. The pressure distribution is determined using measured lattice parameters through 3$^{rd}$-order Birch-Murnaghan equation of state from [19]. Samples were compressed gradually up to ~14-15 GPa at the culet center. In-situ synchrotron XRD in axial diffraction geometry were performed at 16-BM-D beamline at HPCAT at Advanced Photon Source with a wavelength of 0.3100 Å and recorded with Perkin Elmer detector. The measurements were performed along two perpendicular culet diameters (230 μm) in 10 μm steps. The sample thickness (see Table S1) was measured through x-ray intensity absorption using the linear attenuation equation [19]. The diffraction images were converted to unrolled patterns using FIT2D software [21] and then analyzed through Rietveld refinement using MAUD software [22] to obtain the lattice parameters, volume fractions of ω-Zr, microstrains, crystallite sizes, and dislocation density [23] (see supplement).

## 3. Results and Discussion

We assume and then prove that after SPD and phase transformation, the initially isotropic polycrystalline Zr deforms as perfectly plastic and isotropic with a strain path-independent surface of the perfect plasticity $\varphi(s) = \sigma_y(p)$ (Figure 2). To determine the pressure dependence of the yield strength of ω-Zr, the pressure distribution of fully transformed region can be used only, i.e., region around culet center of 3 GPa step and the whole diameters after 3 GPa step. Assuming von Mises yield condition with $\sigma_y = \sigma_y^0 + bp,$ and considering non-hydrostatic stress and heterogeneity along thickness, the equilibrium equation averaged over thickness is advanced to (see supplement):

$$\frac{d\bar{P}}{dr} = -A \frac{\sigma_y^0 + b\bar{P}}{h} \rightarrow \bar{P} = \left(P_0 + \frac{\sigma_y^0}{b}\right) exp\left(-A\, b\, \frac{r-r_0}{h}\right) - \frac{\sigma_y^0}{b}; \quad A = \frac{2(1+0.524b)}{\sqrt{3}(1-0.262b)}, \quad (1)$$

where $P_0$ is the pressure at point $r_0$. From Equation (1),

$$\sigma_y(\bar{P}) = -Ah\frac{d\bar{P}}{dr} = -A\frac{d\bar{P}}{d\left(\frac{r}{h}\right)}. \quad (2)$$

The pressure distributions are plotted vs. $r/h$ in Figure 3. To extract the yield strength utilizing data at all compression steps and positions, pressure distributions from different compression stages are shifted horizontally to the same position. Figure 3 shows that for four different compression stages all pressure distributions overlap with each other and are described by Equation (1) with single pressure dependence $\sigma_y = 1.24 + (0.0965 \pm 0.0016)p$ $(GPa)$. Note that $\sigma_y^0 = 1.24\ GPa$ is converted from the hardness of ω-Zr from [24], $HV$=3.72 GPa, based on the known relationship $\sigma_y^0 = HV/3$, proving that $\tau_y$ is reached with rough-DA. Finite element simulations of the processes in DAC [20, 25, 26] and Figure S1 demonstrate that for different material positions and compression stages, $\varepsilon_p$, $\varepsilon_p^{path}$, and material rotations vary substantially. Consequently, the ability to describe all four curves with single function $\sigma_y(p)$ demonstrates strict proof, for the first time, that for the monotonous loading with rough-DAs, ω-Zr deforms as perfectly plastic and isotropic material with $\varepsilon_p$ and $\varepsilon_p^{path}$-independent surface of perfect plasticity. Since $\varepsilon_p$ and $\varepsilon_p^{path}$ are the only reasons for the strain-induced anisotropy, independence of the yield surface of them implies isotropy also from the theory. Similar, friction stress $\tau_f = \tau_y = \frac{\sigma_y}{\sqrt{3}} = 0.72 + (0.0557 \pm 0.0009)p$ (GPa) is also *independent of* $\varepsilon_p$ *and* $\varepsilon_p^{path}$.

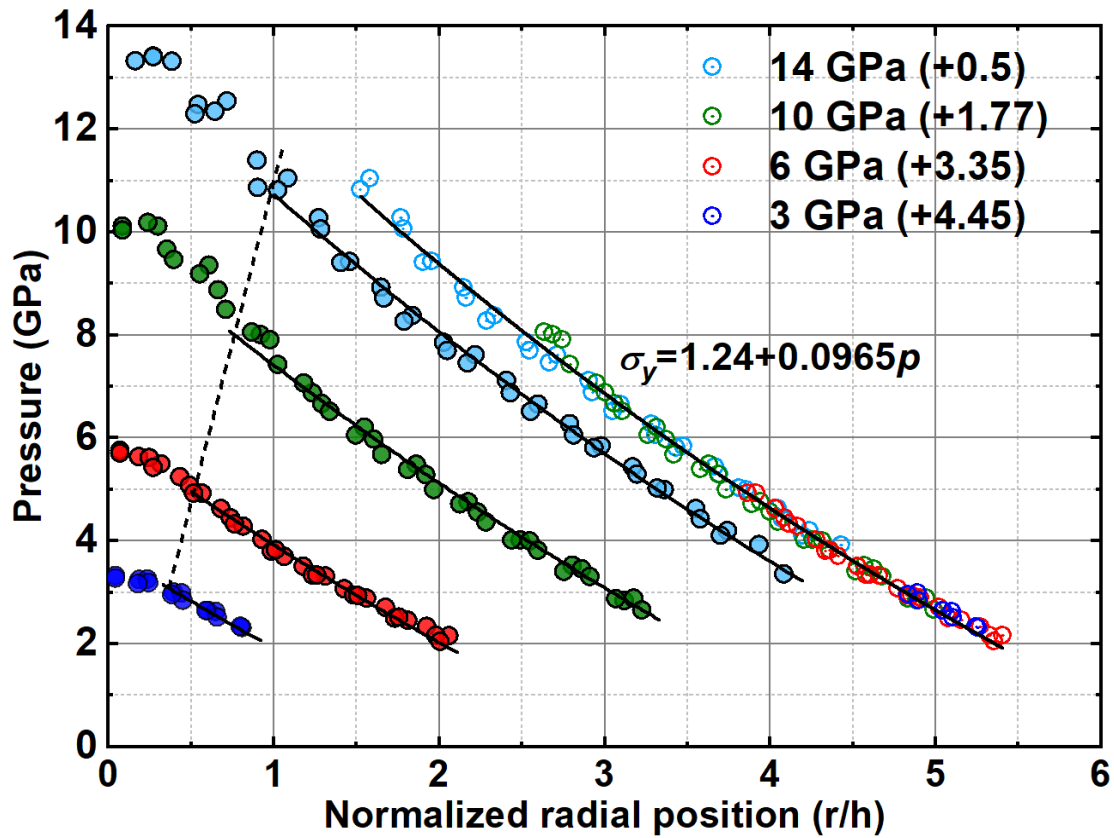

**Figure 3. Pressure in single-phase ω-Zr vs. r/h**. Solid lines correspond to Equation (1) for $\sigma_y^0 = 1.24$ GPa and $b=0.0965$. Equation (1) is not valid around the culet center due to reduction in friction stress to zero at the symmetry axis. Dashed line shows the position where data is truncated. The unified curve for all loadings (necessary for using data from all four compression stages as a single data set) is obtained by shifting each curve (which is allowed by differential Equation (1)) along the horizontal axis by distance shown in parentheses. Note that uncertainty of pressure as well as crystallite size and dislocation density in the following are smaller than the symbols.

We connect perfectly plastic behavior with reaching steady nanostructure. After completing phase transformation in the whole sample, crystallite size $d_\omega$ for 6, 10, and 14 GPa steps scatters between 40 and 60 nm, being practically independent of radius (Figure 4(a)). Dislocation density $\rho_\omega=1.04(19) \times 10^{15}$ m$^{-2}$ is also independent of radius (Figure 4(b)). Since $\varepsilon_p$, $\varepsilon_p^{path}$, and $p$ strongly vary with radius and increasing load, this indicates that steady nanostructure, which is independent of pressure, $\varepsilon_p$, and $\varepsilon_p^{path}$, is reached. Using the general equation for the yield strength as a combination of the Taylor contribution due to dislocation density and Hall-Petch contribution due to grain size [27], we obtain:

$$\bar{\sigma}_y = \tilde{\sigma}_y(p) + \alpha\rho^{0.5} + \beta d^{-0.5}. \qquad (3)$$

Eq. (3) shows consistency between steady states in $\bar{\sigma}_y, \rho,$ and $d$.

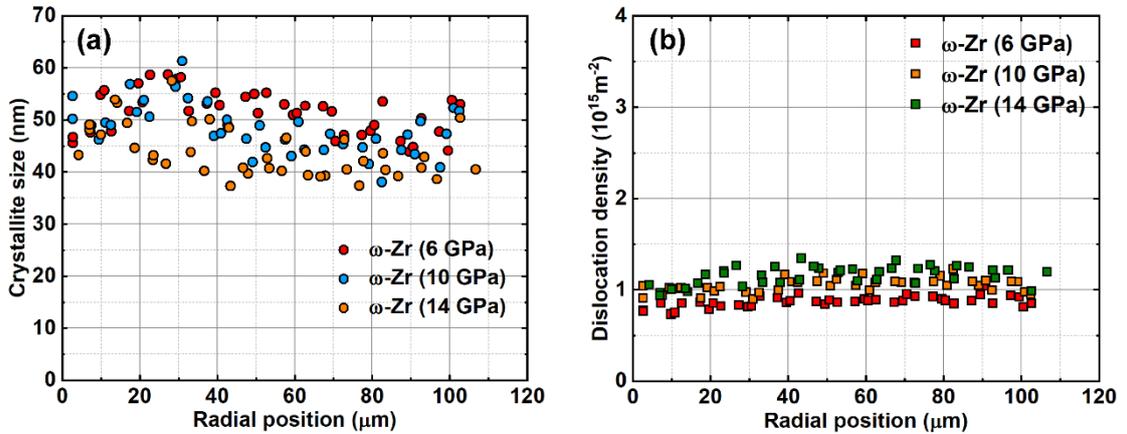

**Figure 4. Radial distribution of the crystallite size (a) and dislocation density (b) in ω-Zr for three loading steps after full transformation.** Since $\varepsilon_p$, $\varepsilon_p^{path}$, and $p$ strongly vary with radius and

increasing load, approximate independence of $d_\omega$ and $\rho_\omega$ of radius and load indicates that steady nanostructure in terms of crystallite size and dislocation density, which is independent of pressure, $\varepsilon_p$, and $\varepsilon_p^{path}$, is reached.

Pressure-independence of the steady microstructure is consistent with pressure-independence of hardness for single-phase Zr for $p < 4$ GPa and $6 < p < 40$ GPa [12], Ti for $p < 4$ GPa and $20 < p < 40$ GPa [28], and Fe for $p < 7$ GPa and $28 < p < 40$ GPa [11]. After HPT of Ni, at the periphery (where the steady state is reached) $0.17 \leq d \leq 0.2$ microns for $3 < p < 9$ GPa [2, 29], which is within an error and is consistent with the pressure-independent hardness for $2 < p < 40$ GPa [11]. Larger grain size for 1 GPa may be related to not reaching a steady state due to smaller friction and plastic strain. Pressure-independent grain size was reached in V [30], Hf, Pt, Ag, Au, Al, Cu, and Cu-30%Zn [11]. The supplement gives some rationales for the pressure independence of the grain size for ω-Zr and difference between known ex-situ and our in-situ rules.

For ω-Zr, with smooth and rough-DA, the steady $\rho_\omega = (0.95 \pm 0.05) \times 10^{15} m^{-2}$ and $(1.04 \pm 0.19) \times 10^{15} m^{-2}$, respectively, and $d_\omega = 49 \pm 1$ nm and $47 \pm 6$ nm, respectively, are practically the same. A completely different situation is with α-Zr, which has three steady states:

1. After multiple rolling at ambient pressure, with $\rho_\alpha = (1.00 \pm 0.02) \times 10^{15} m^{-2}$ and $d_\alpha = 75 \pm 1$ nm.
2. After deformation with smooth anvils, just before initiation of the α-ω phase transformation at 1.36 GPa, with $\rho_\alpha = (1.26 \pm 0.07) \times 10^{15} m^{-2}$ and $d_\alpha = 65 \pm 1$ nm.
3. After deformation with rough-DA, just before initiation of the α-ω transformation at 0.67 GPa, with $\rho_\alpha = (1.83 \pm 0.03) \times 10^{15} m^{-2}$ and $d_\alpha = 48 \pm 2$ nm.

The reason for different steady states cannot be related to the different pressures only because its effect is non-monotonous within a small pressure range. Our results about the existence of multiple steady states are consistent with known results that different ways to produce SPD (e.g., HPT, ECAP, etc.) lead to different steady grain sizes [1-3, 31]. However, different steady dislocation density and crystallite size mean different yield strengths $\sigma_y^i(p)$ (which we could not determine robustly due to the small number of experimental points for single-phase α-Zr) and surfaces of perfect plasticity $\varphi^i(s) = \sigma_y^i(p)$ (Figure 5). Each of these states was obtained at quite different plastic strain and strain paths, so each of them supposed to be independent of $\varepsilon_p$ and $\varepsilon_p^{path}$. But if this is true, how can steady $\rho$, $d$, and $\sigma_y^i(p)$ be different, and which of these steady values should be used in plasticity theory? Thus, the existence of multiple steady states leads to the formulation of a new major challenge in the plasticity and microstructure evolution theories: for which classes of $\varepsilon_p$ and $\varepsilon_p^{path}$ and may be pressure path, material behaves along each of the surfaces $\varphi^i(s) = \sigma_y^i(p)$ with corresponding steady $\rho$ and $d$, and for which loading classes the material behavior jumps from one surface to another with different steady $\rho$ and $d$? When this problem is resolved, one will be able to explain why different SPD technologies lead to different steady $\rho$ and $d$ [1-3, 31], and how to design the loading paths to reduce the $\rho$, and increase $d$ and strength. One of the potential reasons for different steady states may be related to the qualitatively different character of the plastic flow, like transition from the laminar to hierarchical turbulent flow at different scales with different degrees of complexity [32-34].

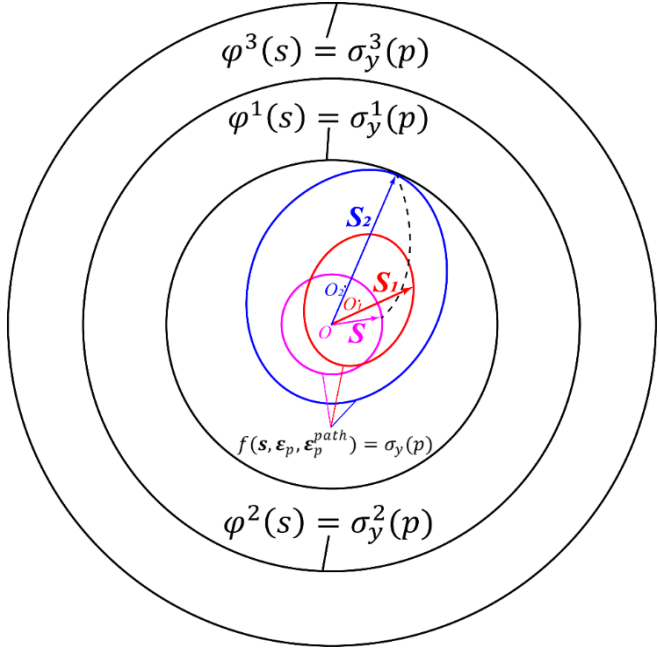

**Figure 5. Evolving yield surface and several fixed surfaces of perfect plasticity.** Part of the schematic with the internal fixed surface of perfect plasticity $\varphi^1(s) = \sigma_y^1(p)$ in "5D" space of deviatoric stresses $s$ at fixed $p$ coincides with that in Figure 2. The difference is in the presence of several other fixed surfaces of perfect plasticity $\varphi^i(s) = \sigma_y^i(p)$ with larger yield strengths $\sigma_y^i(p)$.

## 4. Concluding remarks

In this paper, the *first in-situ* study of the rules of dislocation density, crystallite size, yield surface, and contact friction under high pressure and SPD is presented. In particular, after some critical plastic strain, ω-Zr behaves like perfectly plastic and isotropic, with fixed plastic strain and the strain-path-independent surface of the perfect plasticity $\varphi(s) = \sigma_y(p)$. The perfectly plastic behavior is connected to another rule: crystallite size and dislocation density of α and ω-Zr are getting $p$ (only for ω-Zr), $\varepsilon_p$ and $\varepsilon_p^{path}$-independent and reach steady values. Pressure in single-phase α-Zr is too low to claim pressure independence.

  To provide a robust method to determine $\sigma_y(p)$ and plastic friction stress, we introduce *rough-DA* with increased height of asperities, for which maximum friction $\tau_f = \tau_y$ is reached. We also advanced the simplified equilibrium equation and utilized data after reaching perfectly plastic state only, thus avoiding mixing of the effect of $\varepsilon_p$ and $\varepsilon_p^{path}$ and pressure. That is why the found relationship, $\sigma_y = 1.24 + (0.0965 \pm 0.0016)p$ (GPa), is much more precise than could be obtained with previous methods. Reaching $\tau_f = \tau_y$ implies that the plastic friction $\tau_f = \tau_y = 0.72 + (0.0557 \pm 0.0009)p$ (GPa) is also *independent of* $\varepsilon_p$ *and* $\varepsilon_p^{path}$.

Three different steady states are obtained for α-Zr after multiple rolling and with smooth and rough-DAs, all are independent *of* $\varepsilon_p$ *and* $\varepsilon_p^{path}$. This leads to the new key problem in plasticity theory: for which classes of $\varepsilon_p$, $\varepsilon_p^{path}$, and maybe pressure path material behaves along each of the surfaces $\varphi^i(s) = \sigma_y^i(p)$ and for which loading classes the material behavior jumps from one surface to another? Solution to this problem will allow one to explain why different SPD technologies lead to different steady grain sizes and dislocation densities and how to design the loading paths to reduce the grain size and increase dislocation density and strength. Similar studies can be repeated for any other material system.

Obtained results suggest a more economical way to produce the desired steady nanostructure. Instead of SPD at high pressure, e.g., by HPT, one can reach one of the steady nanostructures by SPD at normal pressure (e.g., by rolling or ECAP) and then reach steady nanostructure with smaller grain size at relatively small plastic strain and low pressure by compression without or with HPT. Rough-DAs also reduce the phase transformation pressure, which will be discussed in future work. Also, rough-DAs eliminate the problem of describing contact friction required for modeling deformational processes in DAC [20, 25, 35]. For traditional HPT with ceramic/metallic anvils, friction reaches the maximum possible level due to large asperities. Utilizing rough-DAs in rotational DAC [36-38] will allow in-situ studies of HPT. In addition, to increase the maximum possible pressure in DAC, toroidal grooves are used [38], which increase friction [18]. This can be done with rough-DAs more uniformly throughout the culet and with smaller stress concentrators.

Importantly, our findings are formulated in the language of plasticity theory (plastic strain and strain path tensors, yield surface, etc.) instead of technological language, which allows one to use the obtained knowledge to significantly enrich fundamental plasticity and the formulation and application of plastic models to various processes.

**Acknowledgements**

The authors thank (a) Drs. Alexander Zhilyaev and María-Teresa Pérez-Prado for providing Zr sample; (b) Dr. Reinhard Boehler for preparing the surface of rough-DA; (c) Drs. Ashraf Bastawros and Bishoy Dawood for the help with asperity measurement. Support from NSF (CMMI-1943710 and MMN-2246991) is greatly appreciated. This work is performed at HPCAT, Advanced Photon Source, Argonne National Laboratory.

**Conflict of interest**

The authors declare no competing interests.

**Data availability**

The data of this study is available from the corresponding authors upon request.

Supplementary Material

# In-situ study of rules of nanostructure evolution, severe plastic deformations, and friction under high pressure

*Feng Lin\*, Valery I. Levitas\*, Krishan K. Pandey, Sorb Yesudhas, and Changyong Park*

**Contents:**

**Supplementary Methods**

1. Evaluation of yield strength under high pressure
2. Derivation of the advanced averaged equilibrium equation
3. Dislocation density estimation
4. Some additional experimental details

**Supplementary Discussion**

1. Scatter in crystallite size and dislocation density in ω-Zr after completing phase transformation
2. The rationale for pressure-independence of the steady grain size
3. Relationship between the yield surface and surface of perfect plasticity
4. Notes on the importance of in-situ studies of severe plastic deformations under high pressure

**Supplementary Figures**

**References**

**Supplementary Methods**

**1. Evaluation of the yield strength under high pressure**

Pressure dependence of the yield strength is of great interest to many disciplines for various reasons. It determines:

(a) strength of structural elements working under extreme loads, in particular, different high-pressure apparatuses, including DAC, rotational DAC, and apparatuses with metallic or ceramic dies for high-pressure torsion;

(b) the maximum pressure that can be achieved in materials compressed in DAC (see Equation (1));

(c) material flow in different technologies, like high-pressure material synthesis, extrusion, forging, cutting, polishing, and ball milling;

(d) maximum possible friction in heavily loaded contacts and related wear;

(e) the level of shear (deviatoric) stresses that can be applied to materials. The shear stresses drastically affect the phase transformations, chemical reactions, and other structural changes [1, 19, 24, 36-38, 40, 41];

(f) plastic flow and geodynamic processes in Earth and other planets, including earthquakes.

There are two approaches to estimate yield strength under pressure in a DAC-like device, which exploit x-ray diffraction in either radial or axial diffraction geometry. With radial diffraction geometry, the yield strength in compression can be estimated from the lattice strains (distortion of crystal lattice planes) measured by synchrotron x-ray diffraction. Since the compression direction is perpendicular to the x-ray beam, lattice strains are detectable because axial compression symmetry and diffraction symmetry do not coincide. With this method, all the components of the elastic strain tensor in single crystals comprising polycrystalline samples can be determined. Combined with high-pressure single-crystal elastic constants, lattice strains can be used to estimate yield strength with proper mechanical assumptions [42]. Despite obtaining a large amount of experimental information and broad usage, this method suffers from several disadvantages:

(a) All measurements are averaged over the diameter of the sample, and the radial gradient of strain and stress fields is unavoidable due to contact friction. The macroscopic stress state also includes shear stresses, which are not included in the treatment. To reduce the effect of friction, a relatively small ratio of the sample diameter to thickness $d/h$ needs to be used, which also limits the axial displacement and applied plastic strain.

(b) When estimating yield strength from the lattice strains, different chosen mechanical assumptions to determine effective elastic properties of the polycrystalline aggregate (Reuss, Voigt, Hill, self-consistent, etc.) leads to different results.

(c) For multiphase materials, lattice strains give an estimation of stress in a single phase only. The mixture theory for the yield strength of multiphase material is not well developed, especially for the large difference in the yield strength of phases [43, 44].

(d) Yield strength depends on the pressure, plastic strain, and grain size that evolve during deformation. By presenting the yield strength versus pressure, all these effects are prescribed to the pressure only, which introduces large errors.

With axial diffraction geometry, yield strength is estimated using radial pressure gradient and sample thickness based on the simplified mechanical equilibrium equation in radial direction $r$ [16-18], combined with the assumption that the friction stress reaches the yield strength in shear $\tau_y$:

$$\frac{d\bar{p}}{dr} = -\frac{2\tau_y(p)}{h}, \tag{S1}$$

where $\bar{p}$ is the pressure, averaged over the sample thickness. Previously, the pressure was measured at the surface using the ruby fluorescence method, and thickness was measured on recovered samples after unloading. Currently, pressure $\bar{p}$ can be measured using x-ray diffraction and thickness using x-ray absorption. The advantage of Equation (S1) is that it does not include constitutive equations and assumptions, making it available for multiphase material. Disadvantages are:

(a) Due to the low friction coefficient of the diamond, the friction stress is much lower than the yield strength in shear $\tau_y$. We found here that for smooth anvils up to 15 $GPa$, the ratio $\tau_f/\tau_y = 0.39 - 0.46$ away from the center characterizes underestimate in the $\sigma_y(p)$ in previous works [16-18]. This is the reason why this method significantly underestimates the yield strength.

(b) Stress $\boldsymbol{\sigma}$ and strain $\boldsymbol{\varepsilon}_p$ tensor fields are strongly heterogeneous along the radius, and material undergoes very different plastic straining path $\boldsymbol{\varepsilon}_p^{path}$ at different positions. Since the yield strength depends on pressure, $\boldsymbol{\varepsilon}_p$, and $\boldsymbol{\varepsilon}_p^{path}$, but is presented as a function of pressure only, this also introduces large errors.

(c) Equation (S1) neglects heterogeneity along the thickness and difference between pressure and normal stresses.

We eliminate all the above drawbacks and advance mechanical equilibrium Equation (S1) to the form of Equation (1) from the main text, which considers the heterogeneity of all stresses across the sample thickness, in the following part.

## 2. Derivation of the advanced averaged equilibrium equation

*Problem formulation.* For compression of a sample in the DAC, $\sigma_{33}$, $\sigma_{11}$, and $\sigma_{22}$ are the normal stress components along the load (vertical), radial, and azimuthal directions, respectively; $\tau_{31}$ is the shear stress; $\sigma_y$ and $\tau_y$ are the yield strength in compression and shear respectively. Compressive stresses are negative. Pressure is defined as:

$$p = -(\sigma_{11} + \sigma_{22} + \sigma_{33})/3 . \tag{S2}$$

All stresses and pressure are functions of $r$ and $2z/h$ in a cylinder coordinate system with the origin at the center of the sample cylinder, where $h$ is the sample thickness; in particular, $p(0)$ corresponds to the symmetry plane $z = 0$ and $p(1)$ corresponds to the contact surface $2z/h = 1$. Pressure (or any stress), averaged over the sample thickness, is defined as:

$$\bar{p} = \frac{1}{h}\int_0^h p\, dz. \tag{S3}$$

The contact friction stress $\tau_f$ is defined by the simplified mechanical equilibrium equation.

$$\frac{d\bar{\sigma}_{11}}{dr} = -\frac{2\tau_f(p(1))}{h} . \tag{S4}$$

The pressure-dependent yield strength in compression $\sigma_y$ and shear $\tau_y = \sigma_y/\sqrt{3}$ (based on the von Mises equivalent stress) are:

$$\sigma_y = \sigma_y^0 + bp; \quad \tau_y = \sigma_y/\sqrt{3} = (\sigma_y^0 + bp)/\sqrt{3} . \tag{S5}$$

Note that $\sigma_y$ depends on the local pressure $p$. At the contact surface, symmetry plane, and for averaged over the thickness, we have different pressures and yield strengths:

$$\sigma_y(1) = \sigma_y^0 + bp(1); \quad \sigma_y(0) = \sigma_y^0 + bp(0); \quad \bar{\sigma}_y = \sigma_y^0 + b\bar{p} . \tag{S6}$$

$$\tau_y(1) = (\sigma_y^0 + bp(1))/\sqrt{3}; \quad \tau_y(0) = (\sigma_y^0 + bp(0))/\sqrt{3}; \quad \bar{\tau}_y = (\sigma_y^0 + b\bar{p})/\sqrt{3} .$$

For maximum possible friction provided by the rough-DA, we have:

$$\tau_f(p(1)) = \tau_y(1) = \frac{1}{\sqrt{3}}\sigma_y(1) = \frac{1}{\sqrt{3}}(\sigma_y^0 + bp(1)) . \tag{S7}$$

With expression in Equation (S7), the equilibrium Equation (S4) specifies as:

$$\frac{d\bar{\sigma}_{11}}{dr} = -\frac{2}{\sqrt{3}}\frac{\sigma_y(1)}{h} = -\frac{2}{\sqrt{3}}\frac{\sigma_y^0 + bp(1)}{h} . \tag{S8}$$

Since we assume that in XRD experiments, the distribution of pressure $\bar{p}(r)$ averaged over the thickness is measured, we need to express $\bar{\sigma}_{11}$ and $p(1)$ in Equation (S11) in terms of $\bar{p}(r)$. Traditionally, this difference is neglected, i.e., it is assumed $\bar{\sigma}_{11} = p(1) = \bar{p}(r)$, which introduces errors.

*Analytical evaluation of the stress and pressure fields.* We assume that the material behaves as perfectly plastic and isotropic macroscopically, with the surface of perfect plasticity $\varphi(s) = \sigma_y(p)$ in the 5D deviatoric stress tensor $s$ space. This surface is independent of the plastic strain tensor $\varepsilon_p$ and its path $\varepsilon_p^{path}$. Such behavior can be achieved after large enough preliminary plastic deformation leading to saturation of hardness [18]. The pressure-dependent von Mises yield condition (i.e., Drucker-Prager yield condition) is assumed:

$$\varphi(s) = \frac{1}{\sqrt{2}}\sqrt{(\sigma_{11} - \sigma_{22})^2 + (\sigma_{11} - \sigma_{33})^2 + (\sigma_{22} - \sigma_{33})^2 + 6\tau_{13}^2} = \sigma_y(p) = \sqrt{3}\tau_y(p). \quad (S9)$$

Equilibrium equations are:

$$\frac{\partial \sigma_{11}}{\partial r} + \frac{\partial \tau_{13}}{\partial z} + \frac{\sigma_{11} - \sigma_{22}}{r} = 0; \quad (S10)$$

$$\frac{\partial \sigma_{33}}{\partial z} + \frac{\partial \tau_{13}}{\partial r} + \frac{\tau_{13}}{r} = 0. \quad (S11)$$

The following assumptions are made:
(a) It approximately follows from the finite element method simulations and DAC experiments: $\sigma_{11} = \sigma_{22}$. Then plasticity condition Equation (S9) simplifies to:
$$(\sigma_{11} - \sigma_{33})^2 + 3\tau_{31}^2 = \sigma_y^2(p) = 3\tau_y^2(p). \quad (S12)$$
(b) Stress $\sigma_{33}$ is independent of $z$. However, it does not mean that:

$$\frac{\partial \tau_{13}}{\partial r} + \frac{\tau_{13}}{r} = 0 \quad \rightarrow \quad \tau_{13} = \tau_0(z)\frac{r_0}{r}. \quad (S13)$$

because at the contact surface, $\tau_0(z)$ may equal constant $\sigma_y$ for all $r$ for material with pressure-independent yield strength. $\sigma_{33}$ that is independent of $z$ means two other terms in Equation (S11) make small contributions to $\sigma_{33}$.

For plane strain, when the term $\frac{\tau_{13}}{r}$ in Equation (S11) is absent, a slightly modified Prandtl's solution for the maximum possible contact friction [45] for stresses that satisfy equilibrium equations and plasticity conditions are:

$$\frac{\sigma_{33}(r)}{\tau_y} = \frac{\sigma_{33}(0)}{\tau_y} + \frac{2r}{h}; \quad (S14)$$

$$\frac{\tau_{13}}{\tau_y} = \frac{2z}{h}; \tag{S15}$$

$$\frac{\sigma_{11}}{\tau_y} = \frac{\sigma_{33}(0)}{\tau_y} + \frac{2r}{h} + \sqrt{3}\sqrt{1-\left(\frac{2z}{h}\right)^2} = \frac{\sigma_{33}(r)}{\tau_y} + \sqrt{3}\sqrt{1-\left(\frac{2z}{h}\right)^2}; \tag{S16}$$

$$\frac{p}{\tau_y} = -\frac{2\sigma_{11}+\sigma_{33}}{3\tau_y} = -\frac{\sigma_{33}(r)}{\tau_y} - \frac{2}{3}\sqrt{3}\sqrt{1-\left(\frac{2z}{h}\right)^2}. \tag{S17}$$

The difference with Prandtl's solution is in multiplier $\sqrt{3}$ instead of 2 in Equation (6) for $\sigma_{11}$. The reason is that we use the von Mises condition and $\sigma_{11} = \sigma_{22}$, which results in Equation (S12), while in Prandtl's solution, the Tresca condition along with plane strain assumption leads to the yield condition $(\sigma_{11} - \sigma_{33})^2 + 4\tau_{31}^2 = \sigma_y^2 = 4\tau_y^2$.

Equation (S16) and Equation (S17) lead to the relationship:

$$\frac{\sigma_{11}}{\tau_y} = -\frac{p}{\tau_y} + \frac{\sqrt{3}}{3}\sqrt{1-\left(\frac{2z}{h}\right)^2}. \tag{S18}$$

Stress $\bar{\sigma}_{11}$ and pressure $\bar{p}$, averaged over the sample thickness are

$$\frac{\bar{\sigma}_{11}}{\tau_y(\bar{p})} = \frac{1}{h}\int_0^h \frac{\sigma_{11}}{\tau_y}dz = \frac{\sigma_{33}(0)}{\tau_y(\bar{p})} + \frac{2r}{h} + \frac{\sqrt{3}\pi}{4} = \frac{\sigma_{33}}{\tau_y(\bar{p})} + \frac{\sqrt{3}\pi}{4}; \tag{S19}$$

$$\frac{\bar{p}}{\tau_y(\bar{p})} = -\frac{\sigma_{33}}{\tau_y(\bar{p})} - \frac{\sqrt{3}\pi}{6}. \tag{S20}$$

We assumed that $\tau_y$ is constant during averaging and then substituted in the result $\tau_y(\bar{p})$. It is possible to avoid this assumption, but the final equations are getting too bulky and unusable analytically for our purposes. Note that the averaged value of $\bar{\sigma}_{11}$ is much closer to the value of $\sigma_{11}(2z/h)$ at the symmetry plane $\sigma_{11}(0)$ than at the contact surface $\sigma_{11}(1)$. For example, $(\sigma_{11}(0) - \sigma_{33})/(\sqrt{3}\tau_y) = 1$, $\sigma_{11}(1) - \sigma_{33} = 0$, and $(\bar{\sigma}_{11} - \sigma_{33})/(\sqrt{3}\tau_y) = 0.79$. Similar, $(p(0) + \sigma_{33})/(2\tau_y/\sqrt{3}) = -1$, $p(1) - \sigma_{33} = 0$, and $(\bar{\sigma}_{11} - \sigma_{33})/(2\tau_y/\sqrt{3}) = -0.79$.

Equation (S19) and Equation (S20) lead to the relationship:

$$\frac{\bar{\sigma}_{11}}{\tau_y(\bar{p})} = -\frac{\bar{p}}{\tau_y(\bar{p})} + \frac{\sqrt{3}\pi}{12}. \tag{S21}$$

We aim to find the relationship between $\bar{\sigma}_{11}$, $\sigma_{11}(0)$, and $\sigma_{11}(1)$. We will use the following identity:

$$\bar{\sigma}_{11} = \sigma_{11}(1)w + \sigma_{11}(0)(1-w); \qquad w := \frac{\bar{\sigma}_{11} - \sigma_{11}(0)}{\sigma_{11}(1) - \sigma_{11}(0)}. \tag{S22}$$

Where $w$ is treated as the weight factor. Utilizing Equation (S16) and Equation (S19), we obtain:

$$w = 1 - \frac{\pi}{4}\frac{\sigma_y(\bar{p})}{\sigma_y(p(0))} = 1 - \frac{\pi}{4}\frac{\sigma_y^0 + b\bar{p}}{\sigma_y^0 + bp(0)}. \tag{S23}$$

Similar,

$$\bar{p} = p(1)w + p(0)(1-w); \qquad w = \frac{\bar{p}-p(0)}{p(1)-p(0)}. \tag{S24}$$

Here we used the same symbol $w$ because, from Equation (S17) and Equation (S20), it has the same expression (Equation (S23)) as for $\sigma_{11}$. Also, we obtain from Equation (S16) and Equation (S18):

$$\sigma_{11}(1) = -p(1) = \sigma_{33}; \quad \sigma_{11}(0) = -p(0) + \frac{\sqrt{3}}{3}\tau_y(p(0)) = -p(0) + \frac{1}{3}\sigma_y(p(0)); \tag{S25}$$

from Equation (S17):

$$p(0) = -\sigma_{33} - 1.155\tau_y(p(0)) = -\sigma_{33} - 0.667\sigma_y(p(0)) = p(1) - 0.667\sigma_y(p(0)); \tag{S26}$$

from Equation (S21):

$$\bar{\sigma}_{11} = -\bar{p} + 0.453\tau_y(\bar{p}) = -\bar{p} + 0.262\sigma_y(\bar{p}) = 0.262\sigma_y^0 + \bar{p}(0.262b - 1). \tag{S27}$$

Elaborating Equation (S26) with allowing for Equation (S6):

$$p(0) = p(1) - 0.667\sigma_y(p(0)) = p(1) - 0.667[\sigma_y^0 + bp(0)] \rightarrow p(0) = \frac{p(1) - 0.667\sigma_y^0}{1 + 0.667b}. \tag{S28}$$

Substitution of Equation (S28) in Equation (S23) and Equation (S24) results in:

$$\bar{p} = p(1)w + \frac{p(1) - 0.667\sigma_y^0}{1 + 0.667b}(1 - w); \qquad w = 1 - (0.785 + 0.524b)\frac{\sigma_y^0 + b\bar{p}}{\sigma_y^0 + bp(1)}. \tag{S29}$$

Resolving linear equations Equation (S29) for $w$ and $p(1)$, we obtain:

$$w = \frac{0.411}{1.910 + b}; \tag{S30}$$

$$p(1) = 0.524\sigma_y^0 + (1 + 0.524b)\bar{p}. \tag{S31}$$

Substituting in Equation (S6) for $\sigma_y(1)$ in Equation (S31), we obtain:

$$\sigma_y(1) = \sigma_y^0 + bp(1) = (\sigma_y^0 + b\bar{p})(1 + 0.524b) \tag{S32}$$

Substituting Equation (S27) and Equation (S32) in Equation (S8) results in the final equilibrium equation for parameters $\sigma_y^0$ and $b$ from the best fit to experiments:

$$\frac{d\bar{p}}{dr} = -\frac{2}{\sqrt{3}}\frac{1 + 0.524b}{1 - 0.262b}\frac{\sigma_y^0 + b\bar{p}}{h}. \tag{S33}$$

Equation (S33) is the final mechanical equilibrium equation expressed in terms of measured pressure $\bar{p}$ averaged of the sample thickness, which is used as Equation (1) in the main text to determine the pressure dependence of the yield strength. It transforms to the known Equation [16-18] for $b = 0$ only. We want to use data from all four compression stages as a single data set. To do this, we must justify a way to combine all data in a single plot. Equation (S33) and its solution in Equation (1) in the main text have the following properties:

(a) Pressure distribution depends on the dimensionless geometric parameter $r/h$ rather than on $r$ and $h$ separately.

(b) Pressure distribution curves for different applied forces and compression can be overlapped by shifting curves along the $r$-axis without changing $\sigma_y(p)$, since change $r \to r + C$ does not violate Equation (S33). Indeed, one can choose the same $p_0$ for all curves and choose constant $C$ for each curve such that $\frac{r+C}{h} = const$ is the same for all curves.

These properties are used in Figure 3 in the main text. Practically, one can choose a fixed $(p_f, r_f)$ point in the $p - r/h$ plane for all curves to pass through. Then the curve that originally passes through the point $(p_f, r_i)$, should be shifted in the positive direction by the distance $(r_f - r_i)/h$, so that the new curve passes through $(p_f, r_f)$. Then we used all the points in the shifted curve in Figure 3 to find the best fit for Equation (S33) (or Equation (1) in the main text).

## 3. Dislocation density estimation

The crystallite sizes and microstrains extracted from the refinement using MAUD were used to estimate the dislocation density. Dislocation density can be expressed as [23]:

$$\rho = \sqrt{\rho_c \rho_{ms}}. \tag{S34}$$

where $\rho_c$ and $\rho_{ms}$ are the contribution to overall dislocation density from crystallite size and microstrain, respectively. Contribution from crystallite size is:

$$\rho_c = \frac{3}{d^2}. \tag{S35}$$

Where $d$ is crystallite size. Contribution from the microstrain is determined by the Equation:

$$\rho_{ms} = k\varepsilon^2/b^2. \tag{S36}$$

Where $\varepsilon$ is the microstrain; $b$ is the magnitude of the Burgers vector; $k = 6\pi A(\frac{E}{G \ln (r/r_0)})$ is a material constant; $E$ and $G$ are Young's modulus and shear modulus, respectively; $A$ is a constant that lies between 2 and $\pi/2$ based on the distribution of strain; $r$ is the radius of crystallite with dislocation; $r_0$ is a chosen integration limit for dislocation core. In this study, $A = \pi/2$ is the gaussian distribution of strain. Moduli $E$, $G$ and their pressure dependence for ω-Zr are taken from [46], respectively. A reasonable value of $\ln (r/r_0)$ being 4 is used [23]. α-Zr has a dominant prismatic slip system of $\{1\bar{1}00\}\langle11\bar{2}0\rangle$ [47-50]. For ω-Zr, a prismatic $\{11\bar{2}0\}\langle1\bar{1}00\rangle$ and basal $\{0001\}\langle1\bar{1}00\rangle$ dominant slip system is suggested based on plasticity modeling [51]. Since the crystal lattice gets compressed under pressure, the length of the Burger vector is calculated using pressure-dependent lattice constants. It is worth to note that when estimating dislocation density using the Williamson-Smallman method, we only consider one dominant dislocation slip system. However, to accommodate arbitrarily imposing plastic strain on polycrystals, auxiliary slip systems are usually needed. With changing orientation of grains during deformation, the Schmid factor of slip systems changes, and thus slip system activities, which is the percentage of plastic strain accommodated by certain slip systems, will be different. This may induce uncertainty in dislocation density estimation. Note that nanocrystals usually do not have a cell structure because cell boundaries are transformed into grain boundaries [4, 52]. That is why the crystallite size is equal to the grain size.

## 4. Some additional experimental details

The material in this study is commercially pure (99.8%) α-Zr (Fe: 330 ppm; Mn: 27 ppm; Hf: 452 ppm; S: <550 ppm; Nd: <500 ppm). We chose Zr as our first test material since Zr and its alloy are widely used in the aerospace, military, medical, and nuclear industries experiencing potential high-pressure environments. The sample thickness during compression (see Table S1) was measured through x-ray intensity absorption using the linear attenuation equation with density corrected to the corresponding pressure, similar to [19].

**Table S1.** The thickness of Zr sample compressed with rough-DAs at corresponding compression steps marked by the peak pressure at the culet center.

| Compression step | initial | 3 GPa | 6 GPa | 10 GPa | 14 GPa |
|---|---|---|---|---|---|
| Thickness (μm) | 163 | 48 | 40 | 32 | 26 |

**Supplementary Discussion**

**1. Scatter in crystallite size and dislocation density in ω-Zr after completing phase transformation**

While the crystallite size and the dislocation density in ω-Zr after completing the phase transformation are independent of the radius (Figure 4), there are some scatters around the average along the radius. Also, the dislocation densities vary slightly between 6, 10, and 14 GPa steps. These scatters cannot be attributed to the dependence of the crystallite size and dislocation density on pressure, plastic strain, and strain path. Indeed, pressure strongly and monotonously reduces, plastic strain strongly and monotonously increases along the radius, and the plastic strain path also changes monotonically. However, there is no clear radial dependence of the crystallite size and the dislocation density. Because of the large fluctuation, the slight difference in the average dislocation density between 6, 10, and 14 GPa steps also cannot be solely attributed to the growing pressure and plastic strain. A possibility is that the observed fluctuations in the crystallite size and the dislocation density after phase transformation completed are due to evolving texture (i.e., dynamically changing distribution of crystallographic orientations and uncharacterized preferred orientations) during the plastic deformation with increasing pressure and errors in post-processing of XRD patterns as described in dislocation density estimation section.

## 2. The rationale for pressure-independence of the steady grain size
### 2.1. Main equations

The existence of steady values of the grain size, dislocation density, and hardness (yield strength) and the parameters they affect are discussed in reviews [1-8]. While pressure dependence of the grain size was not quantitatively analyzed in the literature, some models and correlations are used for the steady grain size at normal pressure. For example, in [53], the following equation is derived:

$$\frac{d_{min}}{b} = A_3 \exp\left(-\frac{\beta Q}{4RT}\right) \left(\frac{D_{PO} G b^2}{\nu_o k T}\right)^{0.25} \left(\frac{\gamma}{Gb}\right)^{0.5} \left(\frac{G}{\sigma}\right)^{1.25}, \quad (S37)$$

where $b$ is the magnitude of Burgers vector, $A_3$ and $\beta$ are constants, $Q$ is the self-diffusion activation energy, $R$ is the gas constant, $k$ is Boltzmann's constant, $T$ is the absolute temperature, $D_{PO}$ is the frequency factor for pipe diffusion, $G$ is the shear modulus, $\nu_o$ is the initial dislocation velocity, $\gamma$ is the stacking fault energy, $\sigma = HV/3$ is the normal stress, and $HV$ is the hardness. Equation (S37) was further transformed in [54] to:

$$\frac{d_{min}}{b} = A_3 \exp\left(-\frac{\beta \lambda T}{4 T_m}\right) \left(\frac{D_{PO} G b^2}{\nu_o k T}\right)^{0.25} \left(\frac{\gamma}{Gb}\right)^{0.5} \left(\frac{G}{\sigma}\right)^{1.25}, \quad (S38)$$

where $T_m$ is the melting temperature, $\beta = 0.037, \lambda = 17.5$. There are also some different simpler relationships in terms of steady hardness $HV$, like:

$$\frac{d_{min}^G}{b} = A \left(\frac{G}{HV}\right)^m \quad (S39)$$

with $m=1.667$ in [54] and $m=1.25$ in [53], and in terms of stacking fault energy $\gamma$, like

$$\frac{d_{min}^\gamma}{b} = C \left(\frac{\gamma}{Gb}\right)^q \quad (S40)$$

with q=0.5 in [53], q=0.4 in [54], and q= 0.653 for HPT and q=0.696 for equal channel angular pressing in [55]. Besides, a relationship in terms of melting temperature is provided in [53, 54]:

$$\frac{d_{min}^{T_m}}{b} = 3818 \exp(-0.00056 T_m). \tag{S41}$$

Note that there is a clear difference between the steady grain size obtained by HPT and processes at normal pressure, like ECAP, which is attributed to the reduction of the minimum grain size with pressure. This is probably related to damage that occurs at low pressure during plastic deformation and stress release, which are not taken into account in the following analysis. Alternatively, or in addition, monotonous straining produces finer grain than the cyclic [3]. For monotonous straining during HPT, larger grain size for low pressure (e.g., for Ni for 1 GPa [2, 29]) may be related to not reaching a steady state due to smaller friction and plastic strain. To be safe, we assume that the following analysis of direct effects of pressure is applicable above some critical pressure, similar to our data for ω-Zr. There are many other parameters that are not taken into account. In particular, we still cannot explain the effect of the deformation with the smooth and rough diamond anvils.

A linear pressure dependence of grain size is assumed:

$$d_{min} = d_{min}^0 (1 - ap), \tag{S42}$$

where $d_{min}^0$ is the grain size at 0 GPa and $a$ is small in comparison with the unity coefficient. We will show in the following evaluations that the pressure-induced reduction in the grain size does not exceed uncertainty in our grain size measurement $6/47 = 0.13$ over a pressure range from 6 to 14 GPa, corresponding to the value of $a$ within:

$$-0.0163 \leq -\frac{0.13}{8} \leq a \leq \frac{0.13}{8} = 0.0163 \ (\text{GPa}^{-1}). \tag{S43}$$

Such a pressure dependence of the grain size is undetectable with our in-situ synchrotron XRD measurements.

*1.2 Pressure dependence of the main material parameters affecting steady grain size*

To analyze the pressure dependence of the grain size, we also assume a linear pressure dependence of any material property $C$:

$$C = C^0 (1 - Ap), \tag{S44}$$

where $p$ is the pressure, $C^0$ is the property at $p=0$, and A is small compared to the unity coefficient. First, we need to collect the pressure dependence of the main properties of ω-Zr that affect it.

Yield strength from this study:

$$\sigma_y = 1.24 + 0.0965p = 1.24(1 + 0.0778p). \tag{S45}$$

Shear modulus from [26]:
$$G = 45.1(1 + 0.0132p). \tag{S46}$$

Bulk modulus from [14]
$$K = 102.4(1 + 0.0286p). \tag{S47}$$

Magnitude of the Burgers vector:
$$b = b_0 \left(1 - \frac{p}{3K}\right) \sim b_0(1 - 0.0033p). \tag{S48}$$

Melting temperature from [56]:
$$T_m = 2125(1 + 0.0094p). \tag{S49}$$

Stacking fault energy (SFE) affects steady grain size, particularly in terms of the combination of twinning and dislocation mechanisms of plasticity. The lower stacking fault energy is the higher contribution of twinning to plastic flow. Significant twinning is observed in α-Zr [48, 50]. It is known that a reduction in grain size suppresses twinning [57]. We are not aware of works reporting twinning in nanocrystalline ω-Zr. Papers [51, 58] quantitatively reproduce experimentally observed in [51] texture by combining different slip modes only. That is why the minimum grain size for ω-Zr may depend on the SFE for reasons other than twinning, like grain size recovery, dislocation absorption by grain boundaries, and dislocation climbing [54].

Let us estimate the pressure-dependence of the SFE by analyzing parameter $a_{sf}$ in $\gamma = \gamma^0(1 + a_{sf}p)$. Since we are unaware of data on the pressure dependence of the SFE for Zr or any other simple hexagonal metal, we will use available data for 9 fcc metals in [59] for the intrinsic stacking faults. Data for the energy of the extrinsic stacking faults are quite similar. Results are collected in Table S2. The largest $a_{sf} = 0.071$ GPa$^{-1}$ is for Ag, which has the smallest $\gamma^0 = 16.9$ mJ/m$^2$, then $a_{sf} = 0.024$ GPa$^{-1}$ for Au with $\gamma^0 = 32.6$ mJ/m$^2$. For all other 7 metals with larger $\gamma^0$, $a_{sf}$ varies between 0.0077 and 0.0186 GPa$^{-1}$. For 3 other fcc metals, Ca, Sr, and Pb, $\gamma$ is getting negative with pressure, i.e., $a_{sf} < 0$. Note that in [60], for Ag $\gamma = 27.3(1 - 0.09p)$ mJ/m$^2$ and for Cu, it is $\gamma = 47.3(1 - 0.01p)$ mJ/m$^2$ in the pressure range $-4$ GPa $\leq p \leq 4$ GPa, i.e., $a_{sf} < 0$ and is close in magnitude to the positive values in [59]. Based on the above results, we assume $a_{sf} = 0.01$ for omega Zr, i.e.,

$$\gamma = \gamma^0 (1 + 0.01p). \tag{S50}$$

*1.3. Evaluating the pressure dependence of the grain size*

While evaluating the effect of pressure in Eqs. (S37)-(S41), we utilize the linear approximation in the Taylor series, e.g.,

$$\frac{(1+bp)(1+cp)^n}{(1+kp)(1+mp)^q} \approx 1+(b-k+cn-mq)p; \qquad (S51)$$

$$\exp(-A(1+ap)) \approx \exp(-A)(1-aAp); \qquad (S52)$$

$$\exp(-A(1+ap))\frac{(1+bp)(1+cp)^n}{(1+kp)(1+mp)^q} \approx e^{-A}(1+(-aA+b-k+cn-mq)p). \qquad (S53)$$

Let us start with Eq. (S40). Using pressure dependence of SFE, $G$ and $b$, we obtain:

$$\left(\frac{\gamma}{Gb}\right)^q \approx \left(\frac{\gamma^0}{G_0 b_0}\right)^q (1+0.0001qp); \qquad (S54)$$

$$d_{min}^\gamma = b\left(\frac{\gamma}{Gb}\right)^q \approx d_{min}^0 (1+(0.0001q-0.0033)p). \qquad (S55)$$

Thus, for any $q$ accepted in the literature, from 0.4 to 0.696, the pressure dependence of $\left(\frac{\gamma}{Gb}\right)^q$ is negligible, and the pressure dependence of $d_{min}^\gamma$ is determined by the pressure dependence of the Burgers vector. For extreme case $q=0$ (which corresponds to the independence of the minimum grain size of $\frac{\gamma}{Gb}$ suggested in [61]), $a = 0.0033 < 0.0163$ (see Eq. (S43)), and the effect of pressure on the grain size according to Eq. (S40) is undetectable experimentally. Since the pressure dependence of $\gamma$ of ω-Zr is not well defined, we will determine the limits of its variation which still make the grain size pressure independent. Assuming $\gamma = \gamma^0(1+a_{sf}p)$, we obtain:

$$d_{min}^\gamma \approx d_{min}^0 (1+(0.0099+a_{sf})qp - 0.0033p). \qquad (S56)$$

For $a_{sf} > 0$, assuming the largest $q=0.696$, we obtain from Eq. (S43) that for $a_{sf} < 0.0381$, the pressure effect on grain size will be undetectable in our experiments. For $a_{sf} < 0$, assuming the smallest $q=0.4$, we obtain from Eq. (S43) that $a_{sf} > -0.0226$. For metals in Table S2 with relatively high SFE like Zr (except for Ag with a significantly lower SFE), their $a_{sf}$ values are within the range of (-0.0226, 0.0381). *Thus, Eq. (S40) agrees with pressure-independent steady grain size in our experiments.*

Next, we evaluate the effect of the pressure dependence of the melting temperature from Eq. (S49) on the $d_{min}^{T_m}$ using Eq. (S41). With the help of Eq. (S52), we obtain

$$d_{min}^{T_m} \approx d_{min}^0 (1-0.0112p). \qquad (S57)$$

Comparison with Eq. (S43) shows that *Eq. (S41) agrees with the pressure-independent steady grain size in our experiments.* Since the dependence of $d_{min}^{T_m}$ in Eq. (S41) comes from the relationship between melting temperature $T_m$ and the activation energy of self-diffusion, we can conclude that the latter also cannot lead to the pressure dependence of the grain size.

The effect of the shear modulus, hardness and the yield strength can be studied based on Eq. (S39). Hardness for Zr is independent of pressure applied during HPT [12], which gives:

$$d_{min}^G \approx d_{min}^0(1 + (0.0132m - 0.0033)p). \quad (S58)$$

For *m*=1.667 in [61], we have $a = -0.0186$, which is slightly larger in magnitude than 0.0163 and is marginally detectable. For *m*=1.25 in [53], one gets $a = -0.0132$, which is undetectable experimentally. For an averaged value $a = -0.0159$, Eq. (S43) is met. Note that in Eq. (S38), the net effect of the shear modulus comes from three terms. If we consider all of them, *m* should be reduced by 0.25. For *m*=1.417, we have $a = -0.0154$ and Eq. (S43) is met. However, initially in [54] for Eqs. (S37) and (S38), combination $(G/\sigma)^{1.25}$ is used instead of $(G/HV)^{1.25}$, where $\sigma$ is the external stress. Then $\sigma$ is equaled to the yield strength, which is not true (because external stress for HPT can be much larger than the yield strength), and then the yield strength is substituted with $HV/3$. Now, we will substitute $HV = 3\sigma_y$ in Eq. (S39) and take pressure dependence of the yield strength into account:

$$\frac{d_{min}^{\sigma_y}}{b} = B\left(\frac{G}{\sigma_y}\right)^m. \quad (S60)$$

We obtain

$$d_{min}^{\sigma_y} \approx d_{min}^0(1 - (0.0646m + 0.0033)p). \quad (S61)$$

Even for smaller *m*=1.25, we obtain $a = 0.0841$, which is more than 5 times larger than the limit in Eq. (S43). Since we do not have any parameter with such a large *a* to compensate for the effect of $\sigma_y(p)$, we can conclude that including pressure-dependent yield strength $\sigma_y(p)$ as one of the parameters affecting the steady grain size contradicts our experimental results, which show pressure-independent minimum grain size. This also excludes the argument that the pressure independent minimum grain size is caused by some specific deformation mechanisms (like dislocations, twinning, or grain boundary sliding) or transition from laminar to turbulent flow [32-34], because all of them are reflected in the experimental pressure dependent yield strength.

Finally, collecting all terms in Eq. (S38) and pressure dependence of $T_m$, $G$, $\gamma$, and $b$, we obtain the combined effect of the pressure on the grain size:

$$d_{min} \approx d_{min}^0(1 + 0.0037p), \quad (S62)$$

which is 4.4 times smaller than can be detected in our experiments. It is clear from Eq. (S54) for *q*=0.5 that the SFE does not contribute to Eq. (S62), i.e., the term with $\gamma$ can be eliminated from Eq. (S38). An increase in the grain size with *p*, while negligible, is counterintuitive. It comes mostly from the term $(G/\sigma)^{1.25}$, which is not well-defined in [53, 54] because $\sigma$ is quite arbitrary. If we eliminate this term, then we obtain:

$$d_{min} \approx d_{min}^0(1 - 0.0128p), \tag{S63}$$

which looks more realistic but is still, according to Eq. (S43), undetectable in our experiment. Thus, the known expressions for the grain size dependence of various material parameters confirm our finding that the steady grain size is pressure independent.

**Table S2.** Stacking fault energy and pressure dependence parameter $a_{sf}$ from [12]

|  | **Co** | **Ni** | **Cu** | **Rh** | **Pd** | **Ag** | **Ir** | **Pt** | **Au** |
|---|---|---|---|---|---|---|---|---|---|
| $\gamma^0$ (mJ/m²) | 168.3 | 153 | 42.4 | 203.4 | 139.5 | 16.9 | 357.2 | 288.1 | 32.6 |
| $a_{sf}$ (GPa⁻¹) | 0.0077 | 0.0078 | 0.0165 | 0.0113 | 0.0186 | 0.0710 | 0.0076 | 0.0111 | 0.0245 |

## 3. Relationship between the yield surface and surface of perfect plasticity

Our results provide the first quantitative proof of the fixed isotropic pressure-dependent surface of perfect plasticity independent of $\boldsymbol{\varepsilon}_p$ and $\boldsymbol{\varepsilon}_p^{path}$, which is far beyond the observation and description in terms of the 'steady hardness'. However, it is well-known that severely deformed materials exhibit plastic strain-induced texture and anisotropy, including the Bauschinger effect described by back stresses. Thus, the traditional yield surface is evolving, anisotropic, and depends on $\boldsymbol{\varepsilon}_p$ and $\boldsymbol{\varepsilon}_p^{path}$ (Figure 2). To resolve this seeming contradiction, we use two different surfaces in "5D" space of deviatoric stresses $\boldsymbol{s}$ at fixed $p$: traditional evolving anisotropic yield surface $f(\boldsymbol{s}, \boldsymbol{\varepsilon}_p, \boldsymbol{\varepsilon}_p^{path}) = \sigma_y(p)$ and fixed isotropic surface of perfect plasticity $\varphi(\boldsymbol{s}) = \sigma_y(p)$. After some critical plastic strain, the yield surface reaches $\varphi(\boldsymbol{s}) = \sigma_y(p)$, and at further monotonous loading, it moves with the deviatoric stress vector s along the fixed isotropic surface $\varphi(\boldsymbol{s}) = \sigma_y(p)$. Thus, the material deforms like perfectly plastic, isotropic with the fixed surface of perfect plasticity. However, during sharp change in loading direction or unloading and reloading in a different direction in the stress space, flow occurs in accordance with actual evolving anisotropic yield surface $f(\boldsymbol{s}, \boldsymbol{\varepsilon}_p, \boldsymbol{\varepsilon}_p^{path}) = \sigma_y(p)$. Due to limited measurement capabilities and strongly heterogeneous fields, and the complexity of equation $f(\boldsymbol{s}, \boldsymbol{\varepsilon}_p, \boldsymbol{\varepsilon}_p^{path}) = \sigma_y(p)$, it is impossible to determine it experimentally. However, finding the surface of perfect plasticity $\varphi(\boldsymbol{s}) = \sigma_y(p)$ is very important because it fully characterizes material's behavior after some critical level of severe plastic deformation and for monotonous loading. Note that the isotropy of the surface of perfect plasticity $\varphi(\boldsymbol{s}) = \sigma_y(p)$ follows not only from experiments but from the theory. Indeed, since initially polycrystalline material with stochastic grain orientation without texture is isotropic, its anisotropy during deformation can come from $\boldsymbol{\varepsilon}_p$ and $\boldsymbol{\varepsilon}_p^{path}$ only, i.e., it is strain-induced. Since $\varphi(\boldsymbol{s}) = \sigma_y(p)$ is independent of $\boldsymbol{\varepsilon}_p$ and $\boldsymbol{\varepsilon}_p^{path}$, the only source for anisotropy disappears. Note that the steady state in the yield strength does not correspond to the steady state in torque in high-pressure torsion [62], mostly due to the complexity of the friction condition. Also, in [63], steady yield strength and dislocation density independent of the changes in strain rate path were obtained in molecular dynamics simulations for a single crystal Ta. These results were called "a tantalizing general hypothesis that merits further scrutiny."

**4. Notes on the importance of in-situ studies of severe plastic deformations under high pressure**

As mentioned in the main text, the effects of severe plastic deformations under high pressure on phase transformations and microstructure evolution are mostly studied with HPT with metallic or ceramic anvils. However, all these results were obtained postmortem after pressure release and further treatment during sample preparation for mechanical and structural studies. The only paper [13] studies the dislocation density and crystallite size in Ni during HPT in a single peripheral region in situ. However, the beam passes also through a significant protrusion part of a sample, which underwent relatively small plastic strain under compression and had lower and very heterogeneous stresses. This brings essential inaccuracy, which varies during the torsion. Also, since data are collected from a single region and for material without phase transformation, the existence of the steady dislocation density and crystallite size can be concluded only. Their independence from pressure and straining path and other our conclusions cannot be drawn from [13]. Note that importance of in-situ molecular dynamics analysis versus ex-situ experiments was stressed in [63].

While we obtained complete phase transformation to ω-Zr in some regions under compression at 3 GPa at the center (Fig. 3) and in the entire sample at 6 GPa, in [12] retaining α-Zr was observed at HPT even after 20 turns. It is written in [12]: "Although this suggests that the complete transformation does not occur, there can be a possibility that a reverse transformation from the ω phase to the α phase might have occurred during mechanical polishing for the preparation of the XRD specimens as reported in an earlier experiment that the reverse transformation occurred during cold machining." A complete transformation in our in-situ experiments confirms the reverse transformation during cold machining and further underlines the importance of in-situ studies. Note that preparing a sample for TEM/SEM may lead to additional changes in the phase fraction, dislocation density, and grain size.

    While our results are consistent with known results [1-8, 11] on the existence of the stationary states after severe plastic straining in terms of hardness, grain size, and dislocation density, and independence of these states of pressure, they mean much more. Our results are obtained directly under pressure versus local pressure at each sample point. Previous results were obtained at the normal pressure and versus averaged pressure over the sample during HPT. Since pressure is distributed very heterogeneously, using an averaged pressure contains a significant error. As an example, independence of the hardness HV=$3\sigma_y^0$ and, consequently, the yield strength of pressure at HPT is obtained for Zr for p<4 GPa and 6<p<40 GPa [12], Ti for p<4 GPa and 20<p<40 GPa [28], for p<4 GPa and 6<p<40 GPa, V [30], Ni, Hf, Pt, Ag, Au, Al, Cu, and Cu-30%Zn [11]. However, it does not imply that the yield strength is independent of the pressure since we obtained explicit pressure dependence for Zr. Similarly, the independence of dislocation density and grain size measured at the ambient pressure of the pressure during HPT does not imply that our in-situ measurements should give independence of dislocation density and grain size of the actual pressure. That is why such independence that we found is a new result. Also, since pressure is distributed very heterogeneously, utilization of an averaged pressure contains a significant error.

**Supplementary Figures**

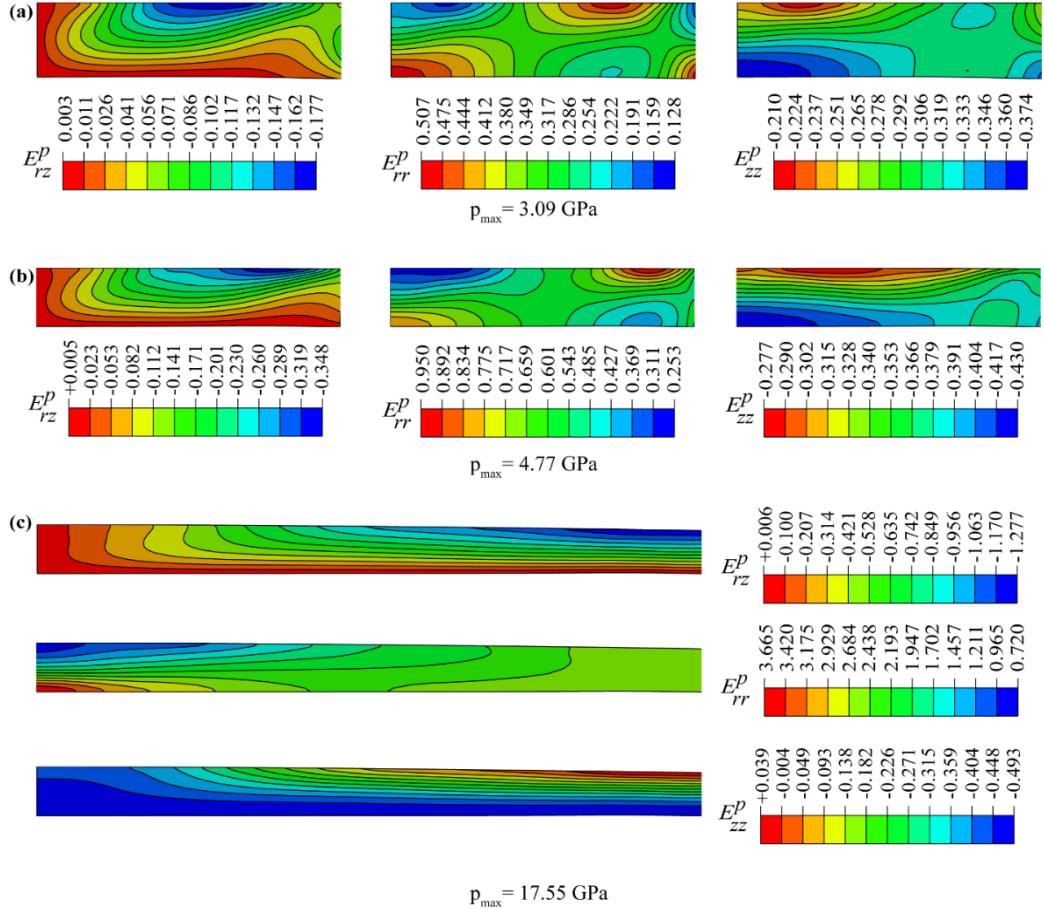

$p_{max}$ = 17.55 GPa

**Figure S1. Distributions of components of Lagrangian plastic strains in a quarter of a sample for three loadings characterized by the maximum pressure in a sample.** Very heterogeneous and nontrivial distributions are observed, caused by heterogeneous contact friction. At the symmetry axis (left side of a sample) and symmetry plane (bottom of a sample), shear strains $E_{rz}^p$ are zero. At the contact surface with a diamond (top of a sample), shear strains and particle rotations reach their maximum due to large contact friction. During compression, each material particle flows radially in the region with larger shear and different proportions of the normal strain, i.e., is subjected to complex nonproportional straining, very different from other particles. Thus, numerous plastic strain tensors and straining paths are realized. Adopted with changes from [26] with permissions.